# STRANGE STARS :
# WHICH IS THE GROUND STATE OF QCD
# AT FINITE BARYON NUMBER ?

Dany PAGE [1] [2]

Astronomy Department, Columbia University

New York, NY 10027, USA

## Abstract

Witten's conjecture about strange quark matter ('Strange Matter') being the ground state of QCD at finite baryon number is presented and stars made of strange matter ('Strange Stars') are compared to neutron stars. The only observable way in which a strange star differs from a neutron star is in its early thermal history and a detailed study of strange star cooling is reported and compared to neutron star cooling. One concludes that future detection of thermal radiation from the compact object produced in the core collapse of SN 1987A could present the first evidence for strange matter.

## 1  Introduction

In a seminal 1984 paper E. Witten[1] proposed that the ground state of baryon matter (or QCD) at finite baryon number might be quark matter and not nuclear matter as always supposed. If the strange quark mass $m_s$ is neglected it is a simple consequence of the Pauli principle that three flavor quark matter has a lower energy than two flavor quark matter. The conclusion usually still holds when $m_s$ is taken into account. (Four flavour is forbidden due to the much larger mass of the charm quark). Quark matter must then be three flavoured, and is called Strange Matter. This simple fact was of course well known long before Witten's paper. What was new is that strange matter may have a lower energy per baryon than nucleon matter. In light of this on can phrase **Witten's conjecture** the following way :
**Two flavor quark matter, at zero pressure, has a higher energy per baryon than the ground state of nucleonic matter ($Fe^{56}$ with $E/A = 930$ MeV) but three flavor quark matter ('strange matter') has a lower energy per baryon.**
For small A Witten's hypothesis is certainly wrong, e.g. for $A = 1$ we obtain for 'strange matter' the $\Lambda$ which has a mass of 1115 Mev, which is higher than the nucleon mass by quite a large amount. For slightly larger A's one can expect nucleonic matter to remain the ground state until some critical value. For heavier nuclei it is easy to understand why they do not decay to the 'true' ground state : one must transform them to strange matter and not to two flavor quark matter, and thus must transform *simultaneously* $A$ quarks $u$ or $d$ into $s$, an $A^{th}$-order weak process. Nuclei are therefore only meta-stable but their lifetime far exceeding the age of the universe.

---

[1] Swiss National Science Foundation Fellow

[2] This is based on preliminary results of work under way in collaboration with Pr J.H. Applegate (Columbia) and Pr M. Prakash (Stony Brook), the final work will be published elsewhere when completed. The content of the present text is the sole responsibility of the speaker (author).

Witten originally proposed strange matter as a candidate for dark matter to be formed in the early universe, but it was then shown that strange matter would not have survived very long and probably would not have formed at all at this stage. Presently, the most probable place where strange matter is expected to exist is in collapsed stars ('neutron stars'). Relativistic heavy ions collisions are also expected to produce quark matter, but not cold (degenerate) quark matter as we are discussing here. We refer to the review by Alcock & Olinto[2] for a general discussion of the subject.

We show here how a star made of strange matter, a 'strange star', can be distinguished from an ordinary neutron star, and will conclude that the 1987A supernova could provide us with a unique oportunity to verify Witten's conjecture.

## 2  A Model for Strange Matter

In order to study quantitatively the properties of strange matter Farhi & Jaffe[3] have proposed a model based on the three following assumptions :

1. The system is well approximated by a degenerate Fermi gas, separated from the vacuum by a phase boundary. The quark phase vacuum thus carry an energy density $B$ (the 'bag constant') of the order of $(100 MeV)^4 - (200 Mev)^4$ (A typical bag model value is $B = (145 MeV)^4 = 57 MeV/fm^3$ ).

2. Quarks are characterized by their current masses :

$$m_u \cong m_d \cong 0 \qquad (1)$$

$$m_s \simeq 100 - 300 MeV \qquad (2)$$

   (The prefered value is $m_s = 175 MeV$).

3. Interactions are modelled using renormalization-group-improved first order QCD, with a QCD coupling constant $\alpha_c \equiv g_c^2/4\pi$.

The model thus depends on three parameters : $B$, $m_s$ and $\alpha_c$. Unfortunately neither of them is yet determined in the particular conditions considered here. However it allows us to test Witten's conjecture within a simple framework.

The calculation starts with the grand thermodynamic potential (Landau's potential) $\Omega(\mu_i)(= -PV)$, which has been calculated by several authors as a function of the various chemical potentials $\mu_i$. The number densities are given as usual by $n_i = -\partial\Omega/\partial\mu_i$, i.e. $n_i = n_i(\mu_i)$. It is found that $\mu_u > p_f(u)$ and $\mu_d > p_f(d)$ ($p_f(i)$ is the Fermi momentum of the $i$ particle), and that at not too high density $\mu_s < (m_s^2 + p_f^2(s))^{1/2}$, reflecting the fact that one gluon exchange is repulsive for relativistic particles and attractive for non-relativistic ones. Thus *interactions favor the appearance of strange quarks*. To find the chemical composition of the system at a given density $n_B$ one must solve the following system of four equations for the four $\mu_i$ $(i = u, d, s, e)$ :

$$\mu_u + \mu_e = \mu_d \tag{3}$$

$$\mu_s = \mu_d \tag{4}$$

$$\frac{2}{3}n_u - \frac{1}{3}n_d - \frac{1}{3}n_s - n_e = 0 \tag{5}$$

$$\frac{1}{3}(n_u + n_d + n_s) = n_B \tag{6}$$

The first two equations impose beta equilibrium, the third charge conservation and the fourth one baryon number conservation. One still must impose the condition of zero pressure :

$$P = -\Omega_u - \Omega_d - \Omega_s - \Omega_e - B = 0 \tag{7}$$

which fixes $n_B$. So, for each set of parameters ($B$, $m_s$ and $\alpha_c$) one obtains a chemical composition and an energy density $\epsilon = \sum(\Omega_i + \mu_i n_i) + B$, and so an energy per baryon $E/A = \epsilon/n_B$ which is compared to the value for nuclear matter. The result of the study is that strange matter does have a lower energy per baryon than nucleonic matter for quite large ranges of parameter values (see Fig.1 in Ref.[3]). Of course the actual values of the parameters can perfectly well lie outside these specific ranges and the model used is questionable, but it makes Witten's idea plausible. The proof or refutation of Witten's conjecture can only come from experiment or observation.

## 3  Strange Stars

Since the most likely place to find strange matter is in collapsed objects ('neutron stars'), the study of strange stars has received quite a lot of attention, in the hope of finding some characteristic feature which would sharply differentiate a strange star from a neutron star. The basic properties of strange stars were first described by Alcock, Farhi & Olinto,[4] and we sumarize here the most relevant ones. The structure of the star, assumed to be spherically symmetric, is obtained by integrating the Oppenheimer-Volkoff equation with the equation of state (E.O.S.), i.e. the relation between $\epsilon$, $n_B$ and $P$, given by the solution of equations 3 to 6 and equation 7 with $P \neq 0$. Unfortunately their global properties, such as mass and radius, turn out to be very similar to those of neutron stars and cannot be used to distinguish them. The (erroneous) announcement of the observation of a very fast rotating object in the remnant of SN1987A with a period of half a millisecond stimulated work on rotational properties of neutron and strange stars, raising the possibility of putting strong constraints on the E.O.S. of hyperdense matter. However the observation was later rejected, ruining a second conceivable discriminant.

We now review the basic structure of neutron stars and strange stars and show that they differ significantly in the crust.

*3.1   Neutron Star Structure : Crust*

A typical neutron star has a mass of 1.4 $M_\odot$ ($M_\odot = 2 \cdot 10^{33} gm$ is the sun mass) and a radius of the order of 10 $km$. Its average density is between $10^{14}$ and $10^{15} gm/cm^3$, and as a zeroth approximation it can be seen as a giant nucleus of $\sim 10^{57}$ nucleons and electrons (and muons) inside it to keep charge neutrality. At closer look it has a crust (mostly solid) made of nuclei on top of the nuclear matter. The uppermost layers are made of material accreted from the interstellar medium, but below this the matter has been processed to nuclear equilibrium, i.e. iron nuclei $Fe^{56}$. At densities above $\rho \sim 10^4 gm/cm^3$ the iron is fully (pressure) ionized and above $\rho \sim 10^6 gm/cm^3$ the electrons become relativistic. This latter fact implies that the (degenerate) electron gas contains a substantial amount of energy and when the density increases still further the electrons have to be absorbed into the nuclei, changing the chemical composition toward more neutron rich nuclei. At a density just below $\rho_{drip} = 4.3 \cdot 10^{11} gm/cm^3$ the equilibrium nucleus is $Kr^{118}$ ($N = 82$, $Z = 36$), but above this point the neutrons produced cannot be all bound into the nuclei and start dripping out : we have now a material of neutron rich-nuclei immersed into a sea of neutrons and electrons. At higher densities the nuclei are deformed, becoming ellipsoidal, more and more elongated, form foils and then start merging together, giving a Swiss cheese-like structure, which dissolves into the nuclear soup around $\rho_{nucl} = 2.8 \cdot 10^{14} gm/cm^3$. We thus have the *outer crust* at densities below $\rho_{drip}$, then the *inner crust* in the neutron drip regime and below them the *core* at densities larger than $\rho_{nucl}$.

*3.2   Strange Star Structure : Crust*

Since strange matter is assumed to be the ground state of baronic matter, anything that touches it will be transformed into strange matter. Thus at first sight strange stars must have a bare quark surface ! This has to be a very sharply defined surface since it is limited by strong forces : at most it is a few fermis thick. Moreover, due to the extremely high density of the surface (several times $10^{14} gm/cm^3$) the electron plasma frequency is so high that thermal radiation cannot be emitted. A bare strange star is a perfect silver sphere, not a black body.

However, this may not be true. The quarks are confined by strong forces, but not the electrons. The electron sea thus extends slightly outside the quark matter, producing a charge disequilibrium at the surface which generates an enormous electric field. Alcock, Farhi & Olinto estimate it to be around $10^{17} - 10^{18} V/cm$. Positively charged ions can sit on top of the quark surface, suspended in the electric field, and form a crust of normal matter hiding the strange matter. The thickness of this crust depends on the strength of the electric field, i.e. on the electron density at the quark surface, which, in Farhi & Jaffe's model of strange matter, in turn depends on the value of the parameters $B$, $m_s$ and $\alpha_c$. There is however a maximum density that this crust can reach, independent of the model : as soon as neutrons drip out of the nuclei they will drop into the strange matter and be swallowed. Consequently, the crust maximum density must be below $\rho_{drip}$ : a strange star can have a crust, but only an outer crust, no inner crust. *This is the crucial difference between a strange star and a neutron star.*

It is not clear how a crust can form on a strange star. Matter falling radially on a bare quark star will have enough kinetic energy to reach the quark surface, but matter coming from an accretion disk could be stopped by the electric field. If a bare strange star is formed in a supernova explosion it is hard to believe that no material will fall back on it. However, once a very thin crust is formed, it will emit radiation probably at supra-Eddington rate and may blow out the crust. This certainly deserves more work and one cannot decide presently whether strange stars have a crust or not.

We will assume in the next section that the strange star we study has a crust, of maximum thickness, i.e. with a density reaching $\rho_{drip}$. The mass of such a crust however is still very small, much less than $10^{-4}$ $M_\odot$.

## 4   Cooling of Strange Stars with a Crust

We consider here the thermal evolution of a strange star *with a crust* and compare it to that of a neutron star. We assume it has been formed hot (supernova) and then cools freely (no reheating from accretion). Its thermal energy is lost by photon emission at the surface and by neutrino emission from the interior. For stars that are not too old the neutrino emission exceeds the photon emission by orders of magnitude. Except when the star is hot (during the first few seconds of its life) the neutrinos produced have a mean free path much larger than the star radius and thus leave the star without any further interaction. Many different neutrino processes can occur in the core of a neutron star or in the strange matter, the most important ones being beta decay with inverse beta decay (URCA process) and variations of it. Table 1 shows the most common processes and their efficiencies : process a) occurs in all neutron stars, the occurence of processes b), c) and d) depends on the model. Process e) of course happens only in quark matter. These emissivities differ by several orders of magnitude and the quark processes are neither more nor less efficient than the nucleon ones, showing that the core neutrino emission cannot allow us to distinguish strange stars from neutron stars. Neutrino emission also occurs in the crust but usually at a less efficient rate making that *during the early phase of the cooling, the core is cooler than the crust*. The surface is also much cooler than the crust due to the poor thermal conductivity of the material below it in the region where the electrons are not yet degenerate. Brown *et. al.*[5] warn that the fast cooling of the core is not seen at the surface until the heat of the crust has had time to diffuse into the core and to the surface. From a simple random walk argument they showed that this diffusion time

$$\tau_d \propto (\text{crust thickness})^2 \qquad (8)$$

and is at least **several decades for a neutron star**. From this one immediately sees the difference between a neutron star and a strange star : since its crust is much thinner the diffusion time will be **a few years for a strange star**. Alcock & Olinto[2] were the first to point this out. This was then confirmed by P. Pizzochero[6] who recently showed how the heat diffusion in the crust can be neatly solved analytically, at the price of some simplifications in the involved physics.

Table 1: Some core neutrino emission processes and their emissivities (in $erg/sec/cm^3$). $Y_e = n_e/n_B$ is the electron fraction and $T_9$ the temperature in units of $10^9$ kelvins.

| . | Process Name | Process | Emissivity $Q_\nu$ |
|---|---|---|---|
| a) | Modified URCA | $n + n' \to n' + p + e^- + \overline{\nu_e}$<br>$n' + p + e^- \to n' + n + \nu_e$ | $\sim 10^{20} \cdot T_9^8$ |
| b) | K-condensate<br>- URCA | $n + K^- \to n + e^- + \overline{\nu_e}$<br>$n + e^- \to n + K^- + \nu_e$ | $\sim 10^{24} \cdot T_9^6$ |
| c) | $\pi$ - condensate<br>- URCA | $n + \pi^- \to n + e^- + \overline{\nu_e}$<br>$n + e^- \to n + \pi^- + \nu_e$ | $\sim 10^{26} \cdot T_9^6$ |
| d) | Direct URCA | $n \to p + e^- + \overline{\nu_e}$<br>$p + e^- \to n + \nu_e$ | $\sim 10^{27} \cdot T_9^6$ |
| e) | Quark URCA | $d \to u + e^- + \overline{\nu_e}$<br>$u + e^- \to d + \nu_e$ | $\sim 10^{27} \alpha_c Y_e^{1/3} T_9^6$ |

*4.1 Detailed Calculation of Strange Star Cooling*

To draw more quantitative conclusions we have studied in detail the heat propagation inside the star using a full stellar evolution code. The basic, general relativistic equations are presented in the review by S. Tsuruta.[7] The code used is an adaptation of the one developed by the author for neutrons stars,[8,9] which includes the best crust physics (thermal conductivity, radiative opacity, neutrino emissivity, specific heat and equation of state) to date. For the quark core, the following physics was used :

- The quark E.O.S. is calculated following Farhi & Jaffe[3] and Alcock, Farhi & Olinto[4] by the method presented in sections 2 and 3. We chose twelve different sets of parameters ($B$, $m_s$ and $\alpha_c$) which span the whole range of values over which strange matter is stable. They are shown in Table 2 together with the properties of the corresponding stars. The crucial quantity obtained here is the electron concentration $Y_e$ because the neutrino emissivity is proportional to $Y_e^{1/3}$. All stars considered have a mass of 1.4 $M_\odot$.

- The specific heat is simply given by relativistic Fermi liquid theory[10] as (for each quark flavor $i$) :

$$C_V = \frac{1}{3}\pi^2 N(0) \cdot T = \frac{k_B^2}{c\hbar^3} p_f^2(i) \left(1 + \frac{8}{3\pi}\alpha_c\right) \cdot T \qquad (9)$$

- The neutrino emissivity [see e) in Table 1] is from Iwamoto's work.[11] The reaction analog to e) with $s$ quark instead of $d$ is also included, but it is only a small correction since it is Cabibbo suppressed.

- The thermal conductivity has been recently evaluated by Haensel & Jerzak,[12] we use their approximate formula :

$$\lambda = 3.4 \cdot 10^{32} \left(\frac{\alpha_c}{0.1}\right)^{-1/2} \left(\frac{n_B}{0.17 fm^{-3}}\right) \frac{1}{T} \quad erg\, cm^{-1}\, s^{-1}\, K^{-1} \qquad (10)$$

*4.2 Results*

The cooling curves of the twelve models are shown (as continuous lines) in Figure 1. One sees that during the first months the surface temperature is the same for all models, this is because during this phase it is completely determined by the upper layers of the outer crust, which are the same for all models. After this phase the temperature drops, corresponding to the cold wave of the core reaching the surface. The time of this phase depends on the thickness of the crust, the thicker the crust the later the temperature drop (see Eq. 8). Figure 2 shows this explicitly with models 6 and 12 : they have the same core cooling rate but different crust thicknesses as can be seen from Table 2. At the end of this period the crust and core are isothermal and we only have a temperature gradient at the surface. The temperature in this phase is entirely determined by the core physics and varies greatly from one model to another (see differences in $Y_e$ in Table 2). Later, photon cooling eventually takes over and the temperature differences disappear (this is not shown in the figure).

Table 2: Strange Stars (1.4 $M_\odot$) Models and Strange Matter Parameters. $R_{star}$ and $R_{core}$ are the star and core radii. During the cooling the envelope contracts slightly, changing the star radius, but the core radius does not vary. $\rho_c$ is the central density. $Y_e = n_e/n_B$ is the electron fraction, given here at the center of the star.

| # | $\alpha_c$ $(g_c^2/4\pi)$ | $B^{1/4}$ $Mev$ | $m_s$ $Mev$ | $R_{star}$ $km$ | $R_{core}$ $km$ | $\rho_c$ | $Y_e \times 10^5$ |
|---|---|---|---|---|---|---|---|
| 1 | 0.3 | 138 | 100 | 11.899 | 11.461 | $6.06 \cdot 10^{14}$ | 0.18 |
| 2 | 0.3 | 138 | 200 | 11.444 | 11.047 | $7.18 \cdot 10^{14}$ | 8.2 |
| 3 | 0.3 | 138 | 300 | 11.493 | 11.091 | $7.49 \cdot 10^{14}$ | 79. |
| 4 | 0.3 | 145 | 100 | 10.999 | 10.640 | $7.96 \cdot 10^{14}$ | 0.09 |
| 5 | 0.3 | 145 | 200 | 10.553 | 10.229 | $9.63 \cdot 10^{14}$ | 4.5 |
| 6 | 0.3 | 150 | 100 | 10.404 | 10.092 | $9.76 \cdot 10^{14}$ | 0.06 |
| 7 | 0.6 | 128 | 100 | 13.327 | 12.750 | $4.16 \cdot 10^{14}$ | 0.02 |
| 8 | 0.6 | 128 | 200 | 12.761 | 12.241 | $4.96 \cdot 10^{14}$ | 2. |
| 9 | 0.6 | 128 | 300 | 12.707 | 12.193 | $5.40 \cdot 10^{14}$ | 35. |
| 10 | 0.6 | 135 | 150 | 12.013 | 11.565 | $5.93 \cdot 10^{14}$ | 0.08 |
| 11 | 0.6 | 135 | 200 | 11.757 | 11.332 | $6.52 \cdot 10^{14}$ | 0.6 |
| 12 | 0.6 | 140 | 100 | 11.621 | 11.208 | $6.55 \cdot 10^{14}$ | 0.003 |

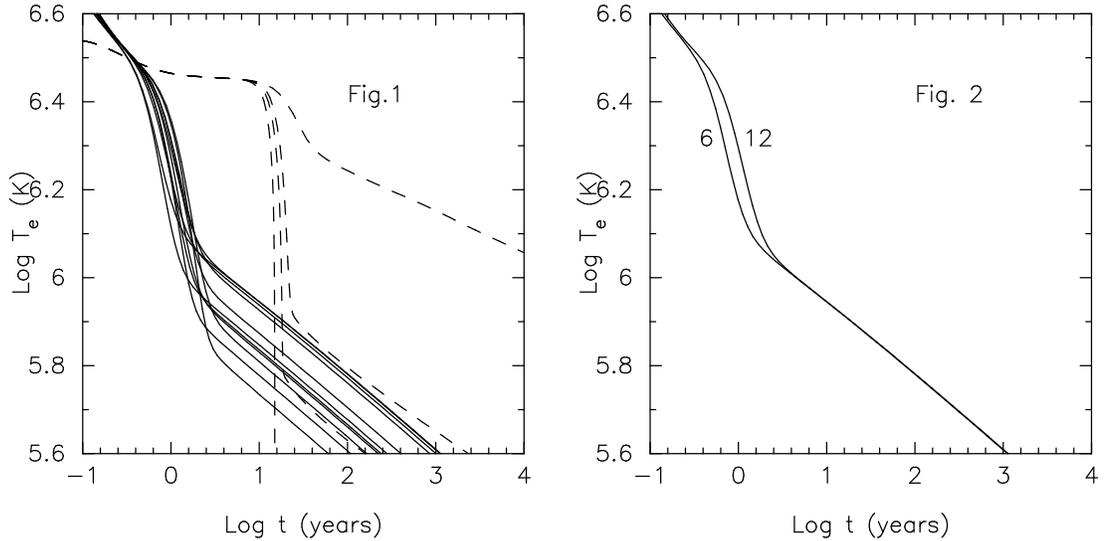

Figure 1: Cooling of Strange Stars vs. Neutron Stars. The continuous curves correspond to the twelve models of strange stars from Table 2, the dashed curves to 1.4 $M_\odot$ neutron stars cooling with the neutrino processes a),b),c) and d) listed in Table 1.

Figure 2: Effect of Crust Thickness on the Cooling : comparison of models 6 and 12.

The dashed lines in Figure 1 are typical cooling curves of neutron stars with the various processes of Table 1. One sees that in the isothermal phase neutron stars can have surface temperatures which are comparable to strange stars and may not be distinguishable from them. However, *all* models of neutron star cooling (see[8,9,13,14] for more models) give a surface temperature above 2 millions kelvins for at least the first decade of the star's life, while all models of strange stars presented here result in a surface temperature below one million kelvins during the same period. *This difference is a characteristic signature of strange stars, due to the thinness of their crusts and the fast cooling of their cores. The compact object recently formed by the core collapse in SN1987A is precisely of this critical age, raising the exciting possibility that detection of thermal radiation from its surface in the near future may unveil the presence of a strange star.*

One must also mention that the Vela pulsar has a temperature estimated[15] of just under one million kelvins for an age of about $10^4$ years, and thus it is well above the cooling curves of strange stars. Moreover A. Alpar[16] warned that strange star models have difficulties in explaining pulsar glitches, observed in particular on the Vela pulsar, providing a second argument for this object not to be a strange star. This does not disprove the strange matter hypothesis since the uncertainty about the possible mechanism which transforms of a neutron star into a strange star perhaps indicates that some of them do not metamorphosize. On the other hand, quark may be superfluid,[17] and a mechanism similar to the one proposed by the author for kaon condensation[8,9] may act here to keep strange stars warmer and could resolve the disprepancy with the Vela pusar. This aspect obviously requires more consideration.

## 5 Conclusion

The strange matter hypothesis of E. Witten claims that strange quark matter is the ground state of QCD at finite baryon number. If this were true it would imply that probably all neutron stars are actually strange stars. The possible masses and radii of strange stars are similar to those of neutron stars, as well as their rotational properties, making them impossible to distinguish from this point of view.

We showed however that, due to the absence of an inner crust, strange stars with a crust have a very particular cooling history during the first decade(s) of their life and have then a much lower surface temperature than neutron stars. In reference to the SN1987A, the detection of thermal radiation from the surface of the compact object then formed, with a temperature below one million kelvins would be a definite signature of a strange star.

**Aknowledgments**

The author benefitted by discussions with J.H. Applegate, P. Pizzochero and M. Prakash. He also wants to thank the Departamento de Fisica of the CINVESTAV and the Instituto de Astronomia of the UNAM for their hospitality during this (1991) summer, as well as the Institute for Nuclear Theory at the University of Washington, where parts of this work were performed.